
\magnification=1200      

\baselineskip=18pt

\hoffset=0.2truecm 

\voffset=0.5truecm 

\hsize= 5.25in 
\vsize=7.5in 

\def\nl{\par\noindent}

\pretolerance=10000

\def\({\c c}
\def\|{\'\i}

\nopagenumbers
\centerline{\bf LIGHT-FRONT QUANTIZATION AND}
\centerline {\bf  SPONTANEOUS SYMMETRY
BREAKING$^{a}$}

\vskip 0.4cm
\centerline {Prem P. Srivastava\footnote{*}{
E-mail:  prem@cbpfsu1.cat.cbpf.br\qquad\qquad {\sl HADRONS94, Gramado, RS,
1994}
\nl {\sl \qquad To be published in:  Proceedings of the
{\it Workshop on Hadron
Physics, 1994}, Gramado, RS, Brasil, World Scientific, Singapore. }  }}

\vskip 0.2cm                                      
\centerline {\sl Instituto de F\|sica, Universidade do Estado do
Rio de Janeiro, RJ}

\vskip 0.4cm

\nl {\bf Abstract:}\quad {\sl The spontaneous symmetry
breaking (and Higgs) mechanism in  the theory
quantized on the light-front ({\it l.f.}), in the
{\it discretized formulation},
is discussed. The infinite volume limit is taken to obtain the {\it continuum
version}. The
hamiltonian formulation is shown to contain a new ingredient
in the form of nonlocal
{\it constraint eqs.} which lead to a {\it nonlocal l.f. Hamiltonian}.
The  description of the broken symmetry here has the same physical
content as in the conventional formulation though arrived at through
a different mechanism. }
\vskip 5truecm 
\vfill
\nl IF-UERJ-0015/94

\eject
\baselineskip=18pt
\pageno=1

{\bf 1.} The  light-front ({\it l.f.})
dynamics of field theories, in which
$\tau=(t+z)\,$ plays the role of evolution parameter, was proposed
by Dirac [1]. The vacuum here is simpler than
in the conventional framework, for example,
in view of the positivity (for massive case) of
the longitudinal momentum $k^+$. It may be very useful to tackle
nonperturbative problems in  QCD [2].

The problems, for example, of the spontaneous symmetry
breaking ({\it SSB}) or the
Higgs mechanism, however, have remained without a clear understanding [2].
We recall that in the usual
equal-time formulation to describe
the {\it SSB} we {\it add} to the theory {\it
external } constraints,  like
$\nabla\phi_{cl}=0$ or that
$\phi_{cl}$ has to minimize the energy functional etc.,
based on physical considerations. On the {\it l.f.} we do
not have such considerations available
not even for minimizing the light-front energy.
It is shown by considering the scalar theory that the {\it l.f.} hamiltonian
formulation contains a new ingredient in the form of {\it nonlocal
constraint equations} which relate the zero modes (bosonic condensates)
with the non-zero modes. These eqns. allow for a description of {\it SSB}
and indicate also that the {\it l.f. hamiltonian is in fact nonlocal and
nonlinear} [3].
The {\it l.f.} dynamics incorporates in it the
necessary constraints in the form of {\it self-consistency conditions}.
The physical results in the {\it l.f.} and the conventional
dynamics coincide, though obtained through different mechanisms.

\bigskip

{\bf 2.} Consider the simpler case of scalar field $\phi$ in two
dimensions with
${\cal L}= \; \lbrack
{\dot\phi}{\phi^\prime}-V(\phi)\rbrack,$.
 Here $\tau\equiv x^{+}=(x^0+x^1)/{\sqrt2}$,
  $x\equiv x^{-}=(x^0-x^1)/{\sqrt2}$, $\partial_{\tau}\phi=\dot\phi ,
	\partial_{x}\phi={\phi}'$, and $d^2x=d\tau dx$. The eq. of motion,
$\,\dot{\phi^\prime}=(-1/2)V'(\phi)\;$, where $ V'$ is
the variational derivative of $V$ with respect to $\,\phi\,$, shows that
$\phi=const. $, is allowed.
We make the separation [3]
$\;\phi(x,\tau)=\omega(\tau)+\varphi(x,\tau)\;$ where $\omega(\tau)$
corresponds to the {\it bosonic condensate} and $\varphi(\tau,x)$ to the
{\it fluctuations} above it. Such a
separation is implied also, we recall, in the discussion of
the degenerate bosonic gas.
The ground state in the quantized theory is  characterized by
the value of $\omega(=\langle 0\vert \phi\vert 0\rangle)$.
In view of the Lorentz invariance requirement we assume [4] for
simplicity that $\omega $ is  independent of $\tau$ so that
${\cal L}={\dot\varphi}{\varphi}^\prime-V(\phi)$ which
describes a constrained
dynamical system. The Dirac procedure [5] will be followed to obtain the
hamiltonian formulation which would permit us
to construct the relativistic and
quantized field theory [1]. The {\it discretrized
formulation} obtained by restricting $x$ to finite size from $\,-L/2\,$ to
$\,L/2\,$ is convenient if we like to avoid using
  generalized functions. The
 {\it physical  limit to the continuum ($L\to\infty\,$)}, however,
  must be taken
to remove the spurious finite volume effects.
Writing the Fourier series expansion

$$\phi(\tau,x) =\omega+{{q_{0}(\tau)}\over\sqrt{L}}+
{1\over\sqrt{L}}\;{{\sum}'_{n, n\ne 0}}\;\;
{q}_n(\tau)\;e^{-ik_n x}\equiv \omega +\varphi(\tau,x)\,\eqno(1)$$

\nl where $\,k_n=n(2\pi/L)$,
$\,n=0,\pm 1,\pm 2, ...\,$
the lagrangian becomes
$\;i{{\sum}_{n}}\;k_n \,{q}_{-n}\;{{\dot q}}_{n}-
                                  \int dx \; V(\phi)$.
The momenta conjugate to ${q}_n$ are
 ${p}_n=ik_n{q}_{-n}$
 and the canonical {\it l.f.} hamiltonian is  $\int \,
 dx \,V(\omega+\varphi(\tau,x))$.
The primary constraints are
${p}_0 \approx 0\;$ and $\;{\Phi}_n \equiv
\;{p}_n-ik_n{q}_{-n}\approx 0\;$ for $\,n\ne 0\,$.
Following the {\it standard} Dirac procedure [5] we establish
three  {\it weak constraints} $\,p_{0}\approx 0$, $\,\beta
\equiv \int dx \,V'(\phi) \approx 0$, and $\,\Phi_{n}\approx 0\,$ for $\,n\ne
0\,$
in the theory and they are shown to be second class. They are taken care of
by defining Dirac bracket.  We find $(\,n,m\ne 0\,)$
$\{{\Phi}_n, {p}_0\}\,=\,0,\qquad
\{{\Phi}_n,{\Phi}_m\}\,=\,-2ik_n \delta_{m+n,0}\;,$
$\quad \{{\Phi}_n,\beta\}\,=\,\{{p}_n,\beta\}\,
=\,-{(1/ \sqrt{L})}\int dx\;\lbrack \,V''(\phi)-V''([{\omega +
q_{0}]/
\sqrt{L}})\,\rbrack \,e^{-ik_nx}\,
\equiv \,-{{\alpha}_{n}/\sqrt{L}},\,$
\quad $\{{p}_0,\beta\}\,=\,-{(1/\sqrt{L})}\int dx\;V''(\phi)\,
\equiv \,-{\alpha/\sqrt{L}},\,$\quad
$\{{p}_0,{p}_0\}\,=\,\{\beta,\beta\,\}\,=\,0\,$.
Implement first the pair of constraints $\;{p}_0\,\approx 0,
\,\beta\,\approx 0$ by defining the star  bracket $\{\}^*$
$\{f,g\}^*\,=\{f,g\}\,-\lbrack\,\{f,{p}_0\}\;
\{\beta,g\}-\,(p_0\,\leftrightarrow
\beta)\rbrack\,({\alpha/\sqrt{L}})^{-1}$.
We may set then ${p}_0\,=0$
and $\beta \,=0$ as {\it strong relations}.
thus removing $\,{p}_0\,$. We find  by inspection that the brackets
$\{\}^*\,$ of the remaining  variables coincide with the standard
Poisson brackets except for the ones involving $q_{0}$ and
$\,p_{n}\,$ ($n\ne0$):\quad
$\,\{q_0,{p}_n\}^*\,=
\{q_0,{\Phi}_n\}^*\,=-({\alpha^{-1}}{\alpha}_n)\,$.
For
$V(\phi)=\,({\lambda/4})\,{(\phi^2-{m^2}/\lambda)}^2\;$,
$\lambda\ge 0, m\ne 0\,$
we find  $\,\{q_0,{p}_n\}^*\,
[\{\,3\lambda\,({\omega+q_{0}/\sqrt{L}})^{2}-m^{2}\,\}L\,+
6\lambda(\omega+q_{0}/{\sqrt L})\int\, dx \varphi +
\,3\lambda \,\int\,dx\,\varphi^{2}\,]=
-\,{3\lambda\,[\,2(\omega+q_{0}/{\sqrt L})\,{\sqrt L}
 q_{-n}
 +\int \,dx\,\varphi^{2}
\,e^{ -ik_{n}x} \,] $.

Next implement the constraints $\,\Phi_{n}\approx 0\,$
($\,n\ne 0$). We have $ C_{nm}\,=\,\{\Phi_{n},\Phi_{m}\}^*\,=
\,-2ik_{n}\delta_{n+m,0}\,$
and its inverse is given by  $\,{C^{-1}}_{nm}\,=\,(1/{2ik_{n}})
\delta_{n+m,0}\;$. The {\it final} Dirac bracket which taking care of all
the constraints is then given by

$$\{f,g\}_{D}\,= \,\{f,g\}^{*}\,-\,{{\sum}'_{n}}\,{1\over {2ik_{n}}}
\{f,\,\Phi_{n}\}^*\,\{\Phi_{-n},\,g\}^*.\,\eqno(2)$$

\noindent where we may now in addition write $\,p_{n}\,=\,ik_{n}
q_{-n}\,$. It is easily shown that
$\{q_0,{q}_0\}_{D}\,=0,
\{q_0,{p}_n\}_{D}\,=\{q_0,\,ik_{n}
q_{-n}\}_{D}\,={1\over 2}\,\{q_0,{p}_n\}^*,
\{q_{n},p_{m}\}_{D}\,={1\over 2}\delta_{nm}$.
Following the well known procedure for taking limit to the continuum:
$\Delta=2\,({\pi/{L}})\to dk\,,\,k_{n}=n\Delta\to k\,,\,\sqrt{L}\,
q_{-n}\to\,lim_{L\to\infty}
 \int_{-L/2}^{L/2}{dx}\,\varphi(x)\,e^{ik_{n}x}\equiv\,\int_{-\infty}^
{\infty}\,dx\, \varphi(x)\,e^{ikx}\,=\,\sqrt{2\pi}{\tilde\varphi(k)}$
for all $\,n $,
$\,\sqrt{2\pi}
\varphi(x)\,=\int_{-\infty}^{\infty}\,dk\,\tilde\varphi(k)\,e^{-ikx}\;$,
and  $(q_{0}/{\sqrt{L}})\to 0\,$.
{}From $\,\{\sqrt{L}q_{m},\sqrt{L}q_{-n}\}_{D}\,=\,
L\,\delta_{nm}/({2ik_{n}})\,$ following from the Dirac bracket between
$q_{m}$ and $p_{n}$ for $n,m\ne 0$
we derive, on using  $\,L\delta_{nm}\to \int_{-\infty}^{\infty}dx e^{i(k-k')x}=
\,{2\pi\delta(k-k')}$, that
$\{\tilde\varphi(k),\tilde\varphi(-k')\}_{D}\,=\,\delta
(k-k')/(2ik)\,$
where  $ k,k'\,\ne 0$.
The  use of the integral representation of the sgn
function leads  to the light-front Dirac bracket
$\{\varphi(x,\tau),\varphi(y,\tau)\}_{D}=-{1\over 4}\epsilon(x-y)$.
It is clear from $\{q_{0},p_{n}\}_{D}$ (or $\{q_{0},\varphi'\}_{D}$)
that the finite volume discretized formulation is hard to
work with.

The expressions of the {\it l.f. hamiltonian} and
the {\it constraint eq.} $\beta=0$ in the continuum take the form

$${ P^{-}\,=\int  dx \,\Bigl [\omega(\lambda\omega^2-m^2)\varphi+
{1\over 2}(3\lambda\omega^2-m^2)\varphi^2+
\lambda\omega\varphi^3+{\lambda\over 4}\varphi^4
\Bigr ]\,}\eqno(3)$$

$$ \eqalign {& \,lim_{L\to\infty}
{1\over L}\int_{L/2}^{L/2} dx \,V'(\phi)\equiv  \cr
& \omega(\lambda\omega^2-m^2)+lim_{L\to\infty} {1\over L}
 \int_{-L/2}^{L/2} dx \Bigl[ \,(3\lambda\omega^2-m^2)\varphi +
 \lambda (3\omega\varphi^2+\varphi^3 ) \,
\Bigr]=0}\,\eqno(4)$$

\nl They may also be obtained readily if we worked directly  in the
continuum formulation. The new
ingredient is the  constraint eq. (4).  Elimination of
$\omega $ would lead to a {\it nonlocal l.f. hamiltonian} in contrast to the
local one in the conventional formulation. This is not
unexpected and does not conflict with the microcausality principle [4].
At the tree level the integrals appearing in (4)
are convergent and on making $L\to\infty$, we find $V'(\omega)=0$ which in the
equal-time formulation is imposed as an  external constraint.

The theory is  quantized through the correspondence
$i\{f,g\}_{D}\to [f,g]$. In the continuum limit we get
$\varphi(x,\tau)= {(1/{\sqrt{2\pi}})}\int dk\;
{\theta(k)}\;
[a(k,\tau)$ $e^{-ikx}+{a^{\dag}}(k,\tau)e^{ikx}]/(\sqrt {2k})$,
were $a(k,\tau)$ and ${a^{\dag}}(k,\tau)$
satisfy the canonical equal-$\tau$ commutation relations,
$[a(k,\tau),{a(k^\prime,\tau)}^{\dag}]=\delta(k-k^\prime)$ etc..
The vacuum state is defined
by  $\,a(k,\tau){\vert vac\rangle}=0\,$,
$k> 0$ and the tree level
description of the {\it SSB} is  given as
follows. The values of $\omega=
\,{\langle\vert \phi\vert\rangle}_{vac}\;$ obtained from
$V'(\omega)=0$
{\it characterize} the different   vacua in the theory.
Distinct  Fock spaces corresponding to different values of $\omega$
are built as usual by applying the creation operators on the corresponding
vacuum state.  The $\omega=0$ corresponds to a {\it symmetric phase}
since the hamiltonian is then symmetric under
$\varphi\to -\varphi$. For $\omega\ne0$ this symmetry is
violated and the  system is
in a {\it broken or asymmetric phase}.
The renormalized constraint eq.
governs the high order corrections to the value
of $\omega$ in the
quantized theory.
It should be stressed that we do {\it not} have
any physical arguments, like for $P^{\pm}$,
in the theory to normal order the constraint
equations.  A factor $L$ may arise in the numerator, for example, in the
$\varphi^{2}$ term which
cancels the $L$ in the denominator in the constraint eq. of the quantized
theory.
A self-consistent {\it l.f. }
hamiltonian formulation can thus be built in the
continuum which also can describe the {\it SSB}.
The extension of the discussion to $3+1$ dimensions and continuous symmetry
is straightforward [3,4].
\bigskip

\nl {\bf References}:

\item{a.} Condensed and adapted from
{\it Spontaneous symmetry breaking mechanism
in light-front quantized field theory}- {\rm Discretized formulation},
{\sl SLAC}/ PPF-9222, April 1992. Available from hep-th@xxx.lanl.gov.
no. 9412193.  {\sl AIP Conference Proceedings 272,
 {\it XXVI Int. Conf. on High Energy Physics}, Dallas, Texas,
August 92,  pg. 2125, ed. J.R. Sanford}.

\item{[1.]}P.A.M. Dirac, Rev. Mod. Phys. {\bf21}, 392 (1949).
\item{[2.]} S.J. Brodsky and H.C. Pauli, {\it Light-Cone}
{\it Quantization of Quantum Chromodynamics}, SLAC-PUB-5558/91; K.G. Wilson,
in {\it Lattice '89}, Proceedings of the
International Symposium, Capri, Italy, 1989, edited by R. Petronzio {\it et
 al.} [Nucl. Phys. B (proc. Suppl.)] {\bf 17} (1990).
 \item{[3.]} P.P. Srivastava, {\it Higgs mechanism
in light-front
quantized field theory},  Ohio State  preprint 92-0012,
{\sl SLAC} data base PPF-9202, December 91; See also Nuovo Cimento
{\bf A 107} (1994) 549 and [4].
\item{[4.]} Lectures on {\it
 Light-front quantized field theory}, Proceedings {\it XIV
 Brazilian National Meeting on Particles and Fields}, Sociedade
 Brasileira de F\|sica, pgs. 154-192, 1993.
 (1993). Available
also from hep-th@xxx.lanl.gov, no.  9312064.
\item{[5.]}P.A.M. Dirac, {\it Lectures
in Quantum Mechanics}, Benjamin, New York, 1964.

\bye